\documentclass[sigconf]{acmart}

\usepackage{float}
\usepackage{booktabs}
\usepackage{multirow, multicol}
\usepackage{subcaption}
\usepackage{xcolor}
\usepackage{diagbox}
\usepackage[most]{tcolorbox}

\usepackage[
type={CC},
modifier={by-sa},
version={4.0},
]{doclicense}
\AtBeginDocument{%
  }

\setcopyright{cc}
\copyrightyear{2026}
\acmYear{2026}
\acmDOI{XXXXXXX.XXXXXXX}
\acmConference[EASE 2026]{The 30th International Conference on Evaluation and Assessment in Software Engineering}{9–12 June, 2026}{Glasgow, Scotland, United Kingdom}
\acmISBN{978-1-4503-XXXX-X/2018/06}

\begin{document}

\title{Evaluating Assurance Cases as Text-Attributed Graphs for Structure and Provenance Analysis}

\author{Fariz Ikhwantri}
\email{fariz@simula.no}
\orcid{0000-0002-7953-0164}
% \author{Dusica Marijan}
% \email{dusica@simula.no}
\affiliation{%
  \institution{Simula Research Laboratory}
  \city{Oslo}
  \country{Norway}
}

\author{Dusica Marijan}
\email{dusica@simula.no}
\affiliation{%
  \institution{Simula Research Laboratory}
  \city{Oslo}
  \country{Oslo}}

%%
%% By default, the full list of authors will be used on the page
%% headers. Often, this list is too long and will overlap
%% The author needs to define a more concise list
% \renewcommand{\shortauthors}{Ikhwantri et al.}

%%
%% The abstract is a summary of the work to be presented in the
\begin{abstract}
An assurance case is a structured argument document that justifies claims about a system’s requirements or properties, which are supported by evidence. In regulated domains, these are crucial for meeting compliance and safety requirements to industry standards.
We propose a graph diagnostic framework for analysing the structure and provenance of assurance cases. We focus on two main tasks: \textbf{(1) link prediction}, to learn and identify connections between argument elements, and \textbf{(2) graph classification}, to differentiate between assurance cases created by a state-of-the-art large language model and those created by humans, aiming to detect bias. We compiled a publicly available dataset of assurance cases, represented as graphs with nodes and edges, supporting both link prediction and provenance analysis. 
Experiments show that graph neural networks (GNNs) achieve strong link prediction performance (ROC-AUC $\approx$0.760) on real assurance cases and generalise well across domains and semi-supervised settings. For provenance detection, GNNs effectively distinguish human-authored from LLM-generated cases (F1 $\approx$ 0.94).
We observed that LLM-generated assurance cases have different hierarchical linking patterns compared to human-authored cases. 
Furthermore, existing GNN explanation methods show only moderate faithfulness, revealing a gap between predicted reasoning and the true argument structure.
\end{abstract}


\begin{CCSXML}
<ccs2012>
   <concept>
       <concept_id>10010147.10010257.10010293.10010294</concept_id>
       <concept_desc>Computing methodologies~Neural networks</concept_desc>
       <concept_significance>500</concept_significance>
       </concept>
   <concept>
       <concept_id>10011007.10010940</concept_id>
       <concept_desc>Software and its engineering~Software organization and properties</concept_desc>
       <concept_significance>300</concept_significance>
       </concept>
   <concept>
       <concept_id>10002951.10003227.10003351</concept_id>
       <concept_desc>Information systems~Data mining</concept_desc>
       <concept_significance>300</concept_significance>
       </concept>

 </ccs2012>
\end{CCSXML}

\ccsdesc[500]{Computing methodologies~Neural networks}
\ccsdesc[300]{Software and its engineering~Software organization and properties}
\ccsdesc[300]{Information systems~Data mining}

% \keywords{Do, Not, Use, This, Code, Put, the, Correct, Terms, for, Your Paper}
\keywords{Knowledge Representation, Evaluation Methodologies, Bias and Safety, Requirements Engineering}

% \received{2 Mar 2026}

%%
%% This command processes the author and affiliation, and title
\maketitle

\section{Introduction}

An assurance case is a set of structured arguments supported by evidence to demonstrate that a system has certain claimed properties in a particular context and environment~\citep{rushby2015understanding}. An assurance case provides confidence in safety, security, or regulatory compliance for a complex system. For example, in aircraft safety~\citep{denney2015dynamic, 10.1109/ICSE.2019.00124}, software security~\citep{weinstock2007arguing, mohamad2021security}, or an Artificial Intelligence (AI) system~\citep{ward2020assurance}. 

\begin{figure}[ht]
    \centering
    \includegraphics[width=0.8\linewidth]{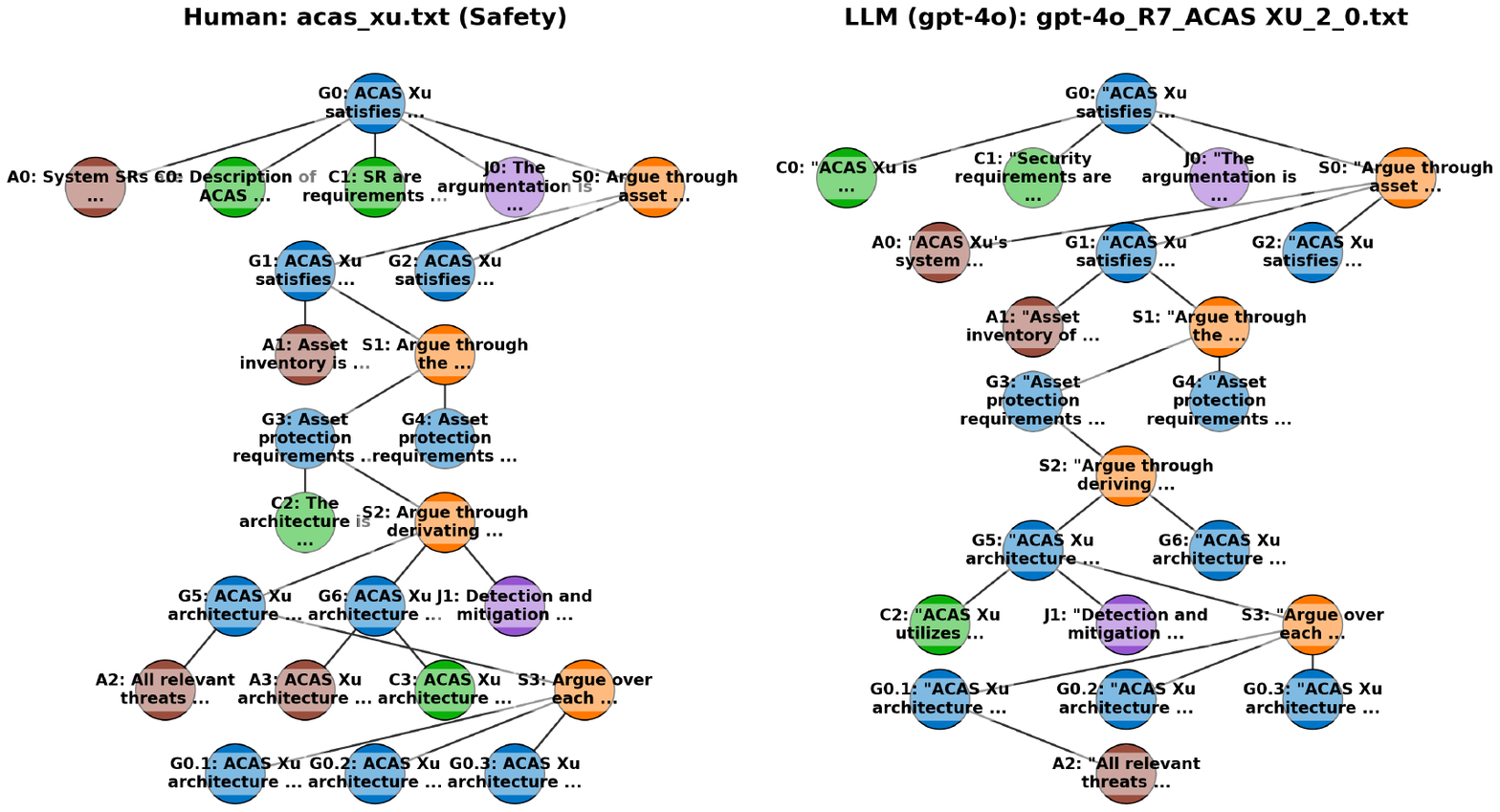}
    \caption{Comparison of human-authored and LLM-generated assurance cases. The LLM-generated case (right) shows a different hierarchical linking pattern and node distribution compared to the human-authored ground truth (left). Different colours represent heterogeneous node types (e.g., Goal, Strategy, Evidence) in the GSN notation. 
    }
    \Description{Comparison of human-authored and LLM-generated assurance cases. The LLM-generated case (right) shows a different hierarchical linking pattern and node distribution compared to the human-authored ground truth (left). Different colours represent heterogeneous node types (e.g., Goal, Strategy, Evidence) in the GSN notation.}
    \label{fig:assurance-case}
\end{figure}

These assurance cases typically comprise a hierarchical argument structure, consisting of claims, supporting arguments, and associated evidence in a tree  structure~\citep{kelly1999arguing, adelard2024} or graphical notation~\citep{kelly2004goal}. Assurance cases are manually constructed and reviewed~\citep{graydon2025examining}. This made them labour-intensive, error-prone~\citep{murugesan2024automating}, and difficult to scale across complex systems~\citep{10.1007/s10270-024-01209-6}.

Automating the analysis of assurance cases becomes increasingly critical as assurance documentation might grow in size and complexity throughout the product life cycle~\citep{denney2015dynamic, Ross2022EngineeringTS}. This automation introduces new challenges for the evaluation and review of assurance cases, particularly in assessing the structure, traceability, and provenance of LLM-generated data. 

Recent studies have increasingly focused on automating the generation of assurance cases using large language models (LLMs)~\citep{sivakumar2024prompting, ODU2025112353, ikhwantri2025explainable}. The first two studies synthesise GSN assurance cases based on predefined templates or safety patterns. The latter generates assurance arguments directly from GDPR for data protection and privacy requirements. 

However, the automatic generation of assurance cases introduces new challenges in trust, traceability, and bias. Automatically generated arguments may appear structurally valid but can lack logical coherence or omit critical evidential links. Moreover, distinguishing between human-authored and LLM-generated assurance cases could be difficult, such as in very large safety requirements. These issues raise concern about reliability and explainability in certification contexts~\citep{graydon2025examining}.

To address these issues, we propose a Textual Attributed Graphs (TAGs)-based modelling approach that leverages Graph Neural Network (GNN) as a representation to reason over and validate the assurance case structures. Unlike previous text-based methods that rely on text similarity~\citep{etezadi2025classificationpromptingcasestudy} or natural language inference~\citep{ikhwantri2025explainable}, our approach models assurance cases as graphs to capture complex inter-dependencies between different elements in an assurance case.

\begin{figure}[ht]
    \centering
    \includegraphics[width=1.0\linewidth]{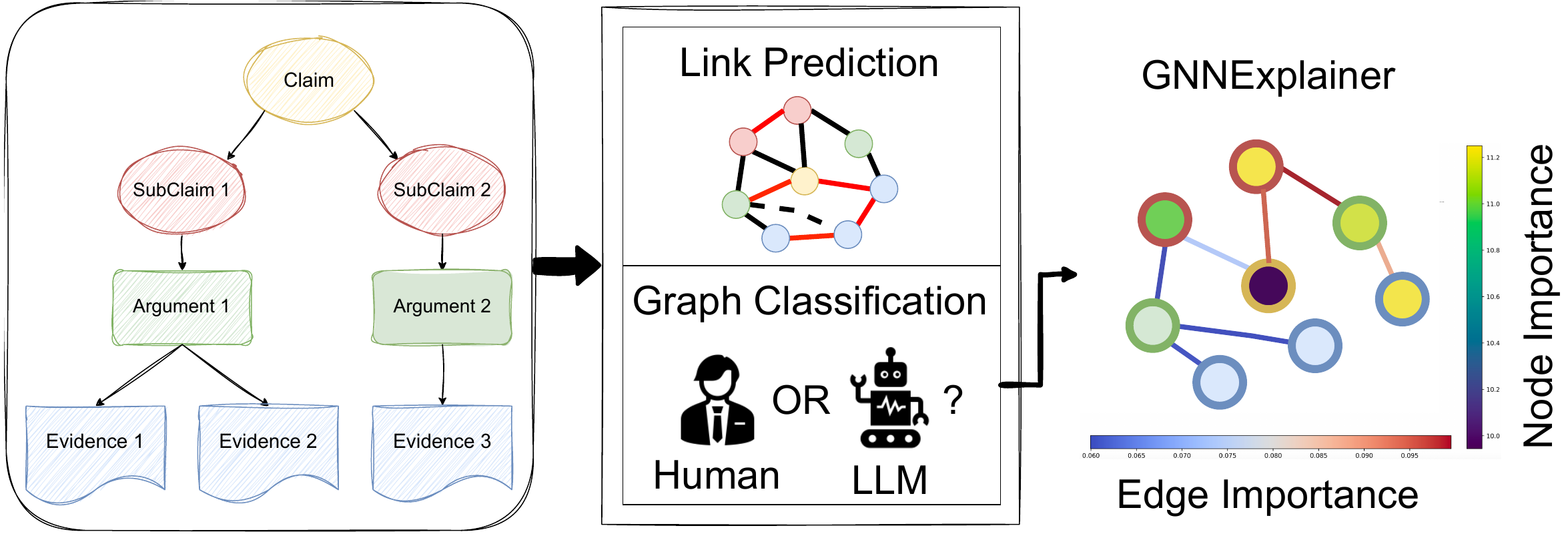}
    \caption{Overview of graph evaluation framework on Assurance Case as text-attributed graphs. Link prediction legend: \textbf{––} correct link, \textcolor{red}{––} incorrect link, and $--$ missing link.}
    \Description{Overview of graph evaluation framework on Assurance Case as text-attributed graphs}
    \label{fig:overview}
\end{figure}

This paper combines a data-centric evaluation of assurance case datasets with an empirical study of AI models applied to structured arguments in Requirements Engineering. Our study makes three primary contributions:
\begin{enumerate}
    \item A curated dataset of human and LLM-generated assurance case graphs. We compile a publicly available assurance case dataset, supporting further research in \textbf{assurance cases in a graph representation and evaluation}.
    \item An empirical evaluation of structural and provenance-related properties of the curated dataset. We focused on two complementary tasks: (1)~\textbf{Link prediction}: infer missing or implicit relationships within assurance case structures; (2)~\textbf{Graph classification}: distinguish between LLM-generated and human ground-truth assurance cases for bias detection.
    \item An analysis pipeline using a graph explainability method for graph-based models to study model behaviour.
\end{enumerate}

Figure~\ref{fig:overview} illustrates our proposed evaluation framework. This framework aims to provide empirical insights into data-centric challenges for evaluating assurance cases, as AI-generated arguments may differ from human-authored ones in structure, connectivity, and provenance. We make the code and data available~\footnote{~\href{https://doi.org/10.5281/zenodo.19816940}{https://doi.org/10.5281/zenodo.19816940}}

\section{Background \& Related Work}\label{sec:rel-work}

\subsection{Assurance Case}

Assurance cases can be represented in different structural formats, which present unique challenges for automated reasoning, scalability, and traceability.

% \subsubsection{Tree Structure}
\paragraph{Tree Structure.}

Early assurance cases were typically modelled using tree structures~\citep{kelly1999arguing, kelly2004goal}, which decompose a top-level claim into subclaims or supporting arguments linked to evidence. This Claim–Argument–Evidence (CAE) model is favoured in safety-critical industries for its simplicity and traceability~\citep{adelard2024}. However, the rigid parent–child hierarchy of trees limits the sharing of nodes, disallows cyclic reasoning, and complicates cross-referencing of assumptions~\citep{murugesan2024automating}. As systems grow in complexity and certification needs evolve, these constraints drive a shift toward graph-based assurance case formats that accommodate richer inter-dependencies and flexible argumentation.

% \subsubsection{Graph Structure}
\paragraph{Graph Structure.}

Goal Structuring Notation (GSN) is a widely used graph-based assurance case notation in industry~\citep{kelly2004goal, adelard2024}. GSN represents the assurance argument using Goals, Strategies, Contexts, Justifications, Assumptions, and Solutions, allowing multiple reasoning paths, reusable argument fragments, and cross-linked contextual information. 
% This expressivity makes GSN particularly suitable for complex systems where arguments may share evidence or assumptions across multiple goals. Despite its advantages, the richer graph structure introduces challenges for automated analysis. 
Safety analysis checking must handle cycles, shared nodes, and varying graphical conventions across organisations, making evaluation more complex compared to tree-based formats~\citep{murugesan2024automating, ikhwantri2025explainable}. These challenges underscore the importance of developing graph-based evaluation and quantitative analysis methods for assurance cases, which is the focus of this study.

Recent studies explore assurance case generation and analysis using natural language processing (NLP) and large language models (LLMs). For example, \citet{murugesan2024automating} proposed goal-directed rule-based and logic programming to automate the semantic analysis of assurance cases. Other studies~\citep{sivakumar2024prompting, ODU2025112353} investigated prompt-based argument generation for safety case templates. Another recent study~\citep{ikhwantri2025explainable} introduces multi-hop reasoning in assurance case structures to model textual semantic relationships.

However, most of these \textbf{previous approaches operate on rule-based or textual representations. These limit their ability to model the structure of the argumentation}. Moreover, explainability and bias detection in automated assurance case evaluation have received limited attention~\citep{ikhwantri2025explainable}. Explainability is important, particularly when the assurance cases are partially or fully generated by LLMs~\citep{graydon2025examining}. 

\subsection{Graph Neural Network (GNN)}

Graph neural networks have demonstrated remarkable success in many domains where relational information is critical, including bioinformatics~\citep{gilmer2017neural, 10.1093/bioinformatics/bty294}, social network analysis~\citep{graphsage10.5555/3294771.3294869, wu2020comprehensive}, recommendation systems~\citep{ying2018graph}, and knowledge graph reasoning~\citep{wang2019kgat}. In recent years, GNNs have been explored in natural language processing (NLP) and reasoning tasks, where graph representations can capture sentence structures, discourse relations, or inter-document dependencies~\citep{yao2019graph, wu2020comprehensive}. In particular, combining GNNs with Large Language Models (LLMs) has shown promise for enhancing structured reasoning, interpretability, and factual consistency~\citep{hu-etal-2025-grag}. Such integration allows leveraging the semantic richness of LLMs while preserving the relational reasoning capabilities of GNNs.

\citet{10697304} categorised three relationships of text and graphs scenarios: (1) pure graphs, (2) text-paired graphs, and (3) text-attributed graphs. In our problem, the assurance case learning on graphs is a type of \textbf{text-attributed graphs} problem. This textual element is the node features $\mathbf{x}_i$, which are typically obtained by encoding textual descriptions (e.g., title and abstract in academic paper networks and claims, strategies, or claim, argument, and evidence in assurance cases) using a pre-trained language model such as BERT~\citep{devlin-etal-2019-bert}. The resulting text embeddings serve as input node features for the GNN, allowing the network to model semantic and structural dependencies within the graph jointly.

This work is related to real-world graph assessment, where the goal is to distinguish real-world topologies from synthetic graphs generated by neural models or classical graph generators. Prior works~\citep{ma2023generated} formalise generated graph detection and show that GNN-based classifiers can reliably identify graphs generated by generative models. \textbf{Our work extends this line of research to a new generator class: large language models that synthesise argumentation graphs rather than structural graphs}.

\section{Methods: Graph-Based Evaluation Framework on Assurance Case}

We consider two primary evaluation tasks: (1) \textit{link prediction}, which captures the model’s ability to learn predicate structure and infer missing or implicit relations between assurance case elements, and (2) \textit{graph classification}, which distinguishes between assurance cases authored by humans and those generated by large language models (LLMs). 
These two tasks are designed to reflect different aspects of assurance case reasoning: structural coherence and provenance bias analysis, respectively. We emphasise that the proposed framework evaluates assurance cases through graph-based analyses. The GNN models are used as \textbf{diagnostic tools to analyse structural patterns rather than to assess assurance case quality directly}.

\subsection{Linking Assurance Case elements}

A \textbf{graph} is formally defined as $\mathcal{G} = (\mathcal{V}, \mathcal{E})$, where $\mathcal{V}$ is the set of nodes (or vertices), and $\mathcal{E} \subseteq \mathcal{V} \times \mathcal{V}$ is the set of edges representing relationships between nodes. Each node $v_i \in \mathcal{V}$ can be associated with an initial features vector $\mathbf{x}_i \in \mathbb{R}^d$ and edge-features vector $\mathbf{e}_{ij}$.

The first task aims to predict whether a link should exist between a pair of nodes $(v_i, v_j)$ in an assurance case as a graph $\mathcal{G}$. 
Formally, given node embeddings $\mathbf{h}_i$ and $\mathbf{h}_j$ learned through a GNN encoder $f_\theta$, the link prediction function estimates the probability of an edge as:
% \begin{equation}
\[
p(e_{ij} \in \mathcal{E}) = \sigma(\mathbf{h}_i^\top \mathbf{W} \mathbf{h}_j)
\]
% \end{equation}

where $\sigma(\cdot)$ denotes the sigmoid activation, and $\mathbf{W}$ is a learnable parameter matrix.
The objective is to maximise the likelihood of existing edges while minimising false predictions for non-existent ones.

This task learns and quantifies connections between elements in the assurance case. By training the model on the \textbf{link prediction task}, we aim to learn recurring structural and semantic dependencies that generalise across different assurance case representations. Specifically, tree-structured safety cases and graph-structured Goal Structuring Notation (GSN) models.

\subsection{Graph Classification: Human vs LLMs-generated Detection}

The second task operates at the graph level and classifies an entire assurance case graph $\mathcal{G}$ as either \textit{human-authored} or \textit{LLM-generated}. 
Given a set of node embeddings $\{\mathbf{h}_i\}_{i=1}^{|\mathcal{V}|}$ learned by the GNN encoder, we apply a graph-level pooling function $\mathrm{READOUT}(\cdot)$, such as mean pooling or attention pooling, to obtain a global representation $\mathbf{h}_G$:
% \begin{equation}
\[
\mathbf{h}_G = \mathrm{READOUT}(\{\mathbf{h}_i\})
\]
% \end{equation}

A classification layer then predicts the label $\hat{y} = \mathrm{softmax}(\mathbf{W}_G \mathbf{h}_G + \mathbf{b}_G)$, where $\mathbf{W}_G$ and $\mathbf{b}_G$ are trainable parameters.

This task aims to detect structural and stylistic biases inherent in LLM-generated assurance cases. The graph classification provides a structure-aware representation to learn provenance detection by distinguishing between human-authored and machine-generated graphs. This provenance detection serves as a diagnostic tool rather than directly assessing the logical validity or regulatory compliance of the case. It can notify safety engineers to automatically generate arguments, flagging them for closer human inspection to ensure they do not contain the hallucinations or generative biases common to LLMs.

Together, these two tasks provide complementary views of assurance cases modelled as text-attributed graphs. Link prediction acts as an auxiliary objective that encourages the model to capture recurring structural and semantic relationships between elements. At the same time, graph classification enables structure-aware provenance detection of an assurance case. Combined, this framework offers a unified perspective for analysing structural characteristics and generation patterns of assurance cases across different structures. 
Importantly, \textbf{the framework is intended to support empirical analysis rather than replace expert review or formal checking}.

\section{Datasets}
We evaluate our proposed framework on three publicly available assurance case datasets collected from prior studies~\citep{10.1109/ICSE.2019.00124, sivakumar2024prompting, ODU2025112353}. The dataset statistics are summarised in Table~\ref{tab:dataset-summary}. These datasets vary in structure, domain, and source of provenance (human vs LLM), covering both Graph structure by Goal Structured Notation (GSN) and Tree structure by Safety Tree format. The first dataset follows a safety tree structure organised \textit{per requirement}, while the last two datasets follow the Goal-Structured Notation format, organised \textit{per product}. This diversity allows us to evaluate the generality of our approach across different assurance case modelling paradigms.

\begin{table}[ht]
\centering
% \small
% \footnotesize
% \scriptsize
\caption{Collections of publicly available Assurance case data from previous studies used in this paper}
\label{tab:dataset-summary}
\begin{tabular}{llrrr}
\toprule
% Dataset & Source & Avg Nodes & Avg Edges & Count \\
Dataset & Src & Avg N & Avg E & Count \\
\midrule
\citet{10.1109/ICSE.2019.00124} & Human & 28.25 & 27.27 & 39 \\
\midrule
\citet{sivakumar2024prompting} & Human & 11.50 & 10.50 & 2 \\
& LLM & 16.15 & 15.10 & 32 \\
\midrule
\citet{ODU2025112353} & Human & 24.20 & 22.40 & 5 \\
& LLM & 17.14 & 16.40 & 185 \\
\bottomrule
\end{tabular}
\end{table}

\subsection{Safety Tree by Requirement}
The safety tree drone dataset from~\citet{10.1109/ICSE.2019.00124} follows a safety tree format, where each graph corresponds to the assurance argument for a single requirement. Unlike the GSN datasets, the linking structure was explicitly defined in the original data, providing an interpretable tree hierarchy among safety claims and supporting evidence. 
This dataset contains only human-authored assurance cases and serves as a complementary benchmark to evaluate model generalisation across different structural representations.

\subsection{GSN Graph by Product}
We compiled these datasets~\citep{ODU2025112353, sivakumar2023gpt4safetycasegeneration} from~\citet{ODU2025112353} and ~\citet{sivakumar2024prompting}. This dataset is in the GSN form. Each graph corresponds to the assurance case of a specific product. 
Each case is composed of textual nodes representing \textit{Goal}, \textit{Context}, \textit{Solution}, \textit{Strategy Argument}, \textit{Justifications}, and \textit{Evidence}. 
For both datasets, we use two variants: human-authored cases and LLM-generated cases produced by GPT-4o~\citep{hurst2024gpt}.

In the original datasets, only the node-level textual content was provided for the LLM-generated cases. 
The linking information (i.e., edges between nodes) was generated in this study using a controlled approach following the complete context prompt with complete contextual information of the context of the Assurace Case, Predicate from the ~\citet{ODU2025112353}. We adjust the prompt to make the same LLM GPT-4o~\footnote{We generated the linking using GPT-4o only, because we did not have access to GPT-4-turbo at the time of the study; this decision was supported by evidence from ~\citet{ODU2025112353} demonstrating higher similarity to human ground-truth.} to infer the predicate structure linking among the argument elements in the generated assurance case. The prompt used for the Linking Prompt for Data Construction with GPT-4o is provided in our replication package.

% (Figure~\ref{fig:link-prompt})

By including both GSN-style and safety tree-style assurance cases with node and edge annotation, our evaluation captures the full range of structural and semantic characteristics found in real-world assurance cases.

\section{LLM-generated Data Quality Analysis}

Before evaluating downstream models, we assess the quality of the LLM-generated~(GPT-4o) assurance cases relative to human-authored cases. In the previous study~\citep{ODU2025112353}, text similarity measures, such as exact string match, BLEU and cosine similarity, were used between all node texts in the ground-truth and all generated node texts by GPT-4o. Since only node-level text, they measured the metrics without considering structure or comparing each node. In this study, we quantify the semantic and structural quality of LLM-generated data using alignment-based metrics adapted from hierarchical evaluation in dependency parsing~\citep{nivre2010dependency}. \textbf{These measures are used solely to characterise dataset properties and are not part of the proposed graph learning framework}.

\subsection{Semantic and Structural Quality Metrics}

\paragraph{Semantic Node Matching.}
Let $H = \{h_1,\ldots,h_n\}$ denote the set of human-authored nodes and
$L = \{l_1,\ldots,l_m\}$ is the set of LLM-generated nodes, where each node is associated with a textual description.
Let $s(h_i,l_j)\in[0,1]$ be a semantic similarity function between node descriptions.

We define a one-to-one semantic matching $M \subseteq H \times L$ as:
\begin{equation}
M = \arg\max_{M'} \sum_{(h,l)\in M'} s(h,l)
% \quad
% \text{s.t. } s(h,l) \ge \tau,
\end{equation}
% where $\tau$ is a similarity threshold.
From the matched node, we can calculate node recall for measuring how many nodes in the LLM-generated nodes are relevant to the human ground-truth. The node recall is defined as:
\begin{equation}
\mathrm{Node\ Recall} = \frac{|M|}{|H|}.
\end{equation}

\paragraph{Edge-Level Structural Matching.}
Since the assurance case is a type of directed acyclic graph, it is difficult to define the graph matching without further study. Hence, we simplify it by computing edge-level matching with the matched nodes.

Let $E_H \subseteq H \times H$ and $E_L \subseteq L \times L$ denote the parent--child
relations in the human ground-truth and LLM-generated assurance cases, respectively. 
Using the matching $M$, we project the human-authored edge set into the LLM node space by retaining only those parent--child pairs for which both endpoints are matched. Each human edge $(h_p, h_c) \in E_H$ with $h_p, h_c \in \mathrm{dom}(M)$ is mapped to the corresponding LLM edge $(M(h_p), M(h_c))$, yielding the projected
edge set $E_H^{M}$.

Similarly, we restrict the LLM-generated edge set to those edges whose parent and child nodes both lie within the range of the matching $M$. The resulting set $E_L^{M}$ therefore contains only relationships between LLM nodes that correspond to matched human nodes. These mappings ensure that both edge sets are defined
over a common node space, which enables structural comparison between the human as ground-truth and LLM-generated assurance cases.

Edge-level precision (\textit{Edge-Prec}), recall (\textit{Edge-Rec}), are computed as:
% and F1-score ()
\begin{equation}
\mathrm{Precision} =
\frac{|E_H^{M} \cap E_L^{M}|}{|E_L^{M}|},
\qquad
\mathrm{Recall} =
\frac{|E_H^{M} \cap E_L^{M}|}{|E_H^{M}|},
\end{equation}
Then, we could compute the Edge F1-score (\textit{Edge-F1}) using $F_1 = \frac{2PR}{P + R}$ from Edge Precision (P) and Edge Recall (R) score.

The edge-level precision, recall, and F1 used here are conceptually similar to attachment-based evaluation metrics in dependency parsing~\citep{nivre2010dependency}, where predicted structural relations are compared against a reference tree. Unlike dependency parsing, human and LLM-generated assurance case node correspondence is not fixed and needs to be established through node matching with text similarity before edge comparison.

\subsection{Semantic and Structural Quality of LLM-generated Data}

\begin{table}[ht]
\small
\centering
\setlength{\tabcolsep}{2.0pt}
\caption{Semantic: Cosine, Node Recall (Node-Rec) and Structural: Edge Precision (Edge-Prec), Recall (Edge-Rec) and F1 (Edge-F1) quality measurement between LLM-generated data from GPT-4o and Human Ground-truth.}
\label{tab:llm-graph-matching}
\begin{tabular}{@{}lccccc@{}}
\toprule
\diagbox{Source}{Metric} & Cosine & Node-Rec & Edge-Prec & Edge-Rec & Edge-F1 \\
% \midrule
% \multicolumn{6}{c}{\citet{sivakumar2024prompting}} \\
\midrule
\citet{sivakumar2024prompting} & .042 & .990 & .197 & .167 & .176 \\
% Std & 0.022 & 0.033 & 0.226 & 0.195 & 0.197 \\
% \midrule
% \multicolumn{6}{c}{\citet{ODU2025112353}} \\
\midrule
% Source & Cosine & Node R & Edge P & Edge R & Edge F1 \\
\citet{ODU2025112353} & .105 & .753 & .489 & .529 & .503 \\
% Std & 0.027 & 0.180 & 0.224 & 0.224 & 0.220 \\
\bottomrule
\end{tabular}
\end{table}

Table~\ref{tab:llm-graph-matching} summarises the semantic and structural alignment between LLM-generated assurance cases and human-authored ground truth. \textbf{These measurements are used solely to characterise LLM-generated part dataset quality and are not part of the proposed learning framework}.

While average cosine similarity scores are low (0.04 to 0.1), average node recall is high (0.75 to 0.9), indicating that LLM-generated data largely recovers the set of human-authored assurance elements at a semantic level. 

In contrast, edge-level precision, recall, and F1 scores vary between the two dataset sources. Notably, the observed edge F1-scores \textbf{could provide a practical upper bound on achievable performance when trained on LLM-generated data only and tested on human ground-truth}, as downstream models cannot recover structural relations that are absent or inconsistent in the underlying data.

These results highlight that LLM-generated assurance cases capture content more reliably than structure, reinforcing the need for structure-aware evaluation in downstream analyses.

\section{Experiments}

\subsection{Research Questions}

\begin{description}
    \item[RQ1.] How well do GNNs learn and generalise link structures in assurance cases across different data provenances?
    \begin{description}
        \item[RQ1a.] How do GNN models perform link prediction on human-authored assurance cases?
        \item[RQ1b.] How do GNN models trained on LLM-generated assurance cases transfer to human-authored cases in cross-provenance link prediction?
        \item[RQ1c.] Does semi-supervised learning that combines human and LLM-generated assurance cases improve GNN robustness and generalisation in link prediction?
    \end{description}
    \item[RQ2] Can graph models distinguish human-authored from GPT-4o-generated assurance cases, and what signals may drive predictions?
    \begin{description}
        \item[RQ2a] How accurately can provenance be classified?
        \item[RQ2b] What do explanation methods suggest about model behavior?
    \end{description}
\end{description}

\begin{table}[ht]
\centering
% \footnotesize
\scriptsize
\caption{Dataset split statistics and graph property. H = Human, M = LLMs}
\label{tab:split_graph}
\begin{tabular}{@{}lrrrrrr@{}}
\toprule
Split & Train & Train & Val & Val & Test & Test \\
Source & H& M & H& M & H& M \\
\midrule
% Avg. \#nodes & 30.433 & 18.557 & 13.000 & 19.920 & 24.364 & 16.705 \\
% Avg. \#edges & 29.467 & 18.142 & 11.000 & 19.180 & 23.455 & 15.410 \\
% Avg\_words & 8.414 & 7.675 & 10.069 & 7.572 & 8.949 & 10.205 \\
% Avg\_chars & 56.472 & 54.094 & 67.880 & 54.410 & 62.707 & 75.075 \\
% Density & 0.134 & 0.062 & 0.123 & 0.057 & 0.057 & 0.063 \\
% Avg\_in\_deg & 0.868 & 0.963 & 0.831 & 0.954 & 0.946 & 0.919 \\
% Avg\_out\_deg & 0.868 & 0.963 & 0.831 & 0.954 & 0.946 & 0.919 \\
% Clustering & 0.000 & 0.006 & 0.000 & 0.001 & 0.000 & 0.023 \\
% Reciprocity & 0.000 & 0.001 & 0.000 & 0.000 & 0.000 & 0.000 \\
% Is\_dag & 1.000 & 0.991 & 1.000 & 1.000 & 1.000 & 1.000 \\
% Homophily & 0.412 & 0.031 & 0.289 & 0.020 & 0.346 & 0.253 \\
% Astv.\_type & -0.144 & -0.507 & -0.404 & -0.483 & -0.074 & -0.393 \\
% Astv.\_degree & -0.520 & -0.487 & -0.600 & -0.437 & -0.322 & -0.426 \\
\#graphs & 30 & 106 & 5 & 50 & 11 & 61 \\
Avg. \#nodes & 30.43 & 18.56 & 13.00 & 19.92 & 24.36 & 16.70 \\
Avg. \#edges & 29.47 & 18.14 & 11.00 & 19.18 & 23.45 & 15.41 \\
Avg. \#words & 8.41 & 7.67 & 10.07 & 7.57 & 8.95 & 10.20 \\
Avg. \#chars & 56.47 & 54.09 & 67.88 & 54.41 & 62.71 & 75.07 \\
Density & 0.13 & 0.06 & 0.12 & 0.06 & 0.06 & 0.06 \\
Avg\_in\_deg & 0.87 & 0.96 & 0.83 & 0.95 & 0.95 & 0.92 \\
Avg\_out\_deg & 0.87 & 0.96 & 0.83 & 0.95 & 0.95 & 0.92 \\
% Clustering & 0.00 & 0.01 & 0.00 & 0.00 & 0.00 & 0.02 \\
% Reciprocity & 0.00 & 0.00 & 0.00 & 0.00 & 0.00 & 0.00 \\ 
% Is\_dag & 1.00 & 0.99 & 1.00 & 1.00 & 1.00 & 1.00 \\
Homophily & 0.41 & 0.03 & 0.29 & 0.02 & 0.35 & 0.25 \\
% Astv\_type & -0.14 & -0.51 & -0.40 & -0.48 & -0.07 & -0.39 \\
% Astv\_degree & -0.52 & -0.49 & -0.60 & -0.44 & -0.32 & -0.43 \\
\bottomrule
\end{tabular}
\end{table}

\subsection{GNN Models}

\textbf{Standard GNN Baselines}. We first employ three widely used GNN architectures—Graph Convolutional Network (GCN)~\citep{kipf2016semi}, Graph Attention Network (GAT)~\citep{velivckovic2017graph}, and GraphSAGE~\citep{graphsage10.5555/3294771.3294869}. These serve as our primary baselines to establish how well GNN neighbourhood aggregation and attention mechanisms capture the local structural and semantic relations within assurance graphs.

\textbf{Graph Foundation Model (UniGraph)}. To address the challenges of \textbf{data scarcity and cross-domain generalisation inherent in our dataset}, we incorporate UniGraph~\citep{10.1145/3690624.3709277}. As a state-of-the-art graph foundation model, UniGraph unifies masked language models and GNNs within a single learning algorithm through self-supervised pre-training (\textbf{MLM Self-Supervised+GNN}). The prior study shows it is capable of few-shot learning. This capability allows the model to extract complex structural semantics effectively without requiring massive, multi-source training datasets, making it ideal for our proof-of-concept evaluation.

\textbf{Few-Shot Graph Classification (GraphPrompt)}. For the graph classification task, we additionally evaluate GraphPrompt~\citep{10.1145/3543507.3583386}, a recent framework that adapts the link prediction task for pre-training graph models. The pre-trained models could be used in downstream tasks via soft-prompt. This soft-prompt enables few-shot inference where the human-authored ground-truth dataset is limited in size and naturally imbalanced compared to the LLM-generated data. We initialise GraphPrompt using the best-performing weights from the link prediction pre-training, enabling the model to maximise structural knowledge transfer for provenance detection.

All graph-based models in the experiment use BERT-base-uncased, which are fixed with no gradient update except the UniGraph model. We use HuggingFace Transformer~\citep{wolf-etal-2020-transformers} and Pytorch Geometric~\citep{fey2025pyg} to implement experiments for this study.

\paragraph{Hyperparameters} 

We set the learning rate to 1e-5, the weight decay to 0.01, and 200 epochs. The GCN/GAT/SAGE baselines use hidden size 256; “-3 layers” denotes three stacked GNN layers (the base variants use a single layer). GAT uses one attention head. UniGraph follows its default configuration with hidden size 768, two layers, eight heads, dropout 0.1, $\lambda$=0.1, and mask\_rate=0.30. The batch sizes are one for the baselines and four for UniGraph.

\subsection{Text-Only Models}

% \paragraph{LLM for Link Prediction.}

\textbf{Decoder LLM} We evaluate LLMs as link predictors on assurance case graphs. Similarly, with GNN for assurance case training and testing, we build a balanced set of candidate edges by taking all directed parent→child links as positives and sampling an equal number of non‑edges as negatives. Each candidate pair is converted to a short instruction prompt asking whether a link exists between the two node descriptions and requiring a Yes/No answer. To obtain a probability, the LLM is queried five times with stochastic decoding. Each generation is parsed into a binary vote; the predicted link probability for a pair is the empirical frequency of “Yes” across samples. This frequency also serves as an uncertainty indicator (values near 0.5 denote ambiguity).

To perform in‑context learning (ICL), we include labelled examples of assurance case graphs from the training set with balanced positives/negatives that exclude the queried nodes. The procedure is model‑agnostic and works with any decoder LLM. In this study, we use Llama 3 models (3.2-1B and 3.1-8B)~\citep{grattafiori2024llama} and Qwen2.5-7B~\citep{qwen2}. We set the models to take 6 in-context pairs of node examples from the training set pool, which consists of 3 positives and 3 negatives in the link prediction task. In the graph classification task, we set the model to take 4 graph examples, which consist of 2 human and 2 llm sample from the training set pool.

For the decoder LLM baseline, we explicitly evaluate uncertainty arising from both stochastic decoding and prompt conditioning. We perform five independent sampling runs per instance to capture inference-time stochasticity. In addition, as fine-tuning or re-training LLMs with different model seeds is computationally prohibitive, we approximate model-level variability by randomly resampling the in-context examples. For each prediction, we record the sampled outputs, the selected in-context examples, and the random seed as part of the released evaluation artefact. Full prompt templates, example instantiations, and generation scripts are released as part of the public replication package.

\subsection{Metrics}
\label{sec:metrics}

To evaluate the proposed framework, we consider metrics for both \textbf{link prediction}, \textbf{graph classification}, as well as explanation faithfulness for the graph classification.

% \textit{Precision}, \textit{Recall},
For \textbf{link prediction}, we report area under the Receiver Operating Characteristic Curve (\textit{ROC-AUC}) and F1-score (\textit{F1}). We compute precision, recall, and F1 with 0.5 threshold probabilities. 
These metrics measure the model’s ability to infer valid and missing links between assurance case elements. These metrics are computed by comparing predicted links against verified human-annotated connections. 
% Performance is reported over all graphs using ROC-AUC. 

For \textbf{graph classification}, we evaluate the model’s capacity to distinguish between human-authored and LLM-generated assurance cases using standard metrics: Accuracy (\textit{Acc}) and F1-score (\textit{F1}). High classification performance indicates that the model captures structural or semantic biases between human and LLM-generated reasoning patterns.

To assess \textbf{explainability} of learned graph classification models, we use the graph fidelity (\textit{Fid}) metrics following \citet{amara2024graphframexsystematicevaluationexplainability} and graph explanation faithfulness~(\textit{GEF})~\citep{agarwal2023evaluating} from PyTorch Geometric by using GNNExplainer~\citep{ying2019gnnexplainer}. 
GEF actually measures unfaithfulness as the deviation between the model’s output change and the masked subgraph’s predicted score. Fidelity quantifies the consistency between the explanation and the model’s decision.

\section{Results}

% \subsection{Link Prediction}
% \subsection{RQ1 (Human $\rightarrow$ Human): Gold Baseline}
\subsection{RQ1: Link Prediction}
\paragraph{RQ1a: Gold Standard Link Prediction}

% \input{tables/link_pred_r1}

% Table~\ref{tab:r1-baseline} 
% Table~\ref{tab:combined-all-scenarios} presents the performance of various baseline and graph-based models when trained and evaluated on human-authored assurance cases, representing the gold-standard reasoning structures. The results demonstrate that incorporating graph structure substantially improves link prediction performance over text-only encoders.
Table~\ref{tab:combined-all-scenarios} presents the performance of various baseline and graph-based models evaluated on human-authored assurance cases. \textbf{GNN models trained and tested on human-authored assurance cases achieve strong performance, with several architectures, especially SAGE and UniGraph, substantially outperforming text-only decoder LLM baselines}. Across the Human → Human setting, GNN-based models achieve higher AUC than Decoder LLM, with SAGE (.796 AUC, .816 F1) and UniGraph (.848 AUC, .739 F1), except for GCN. This indicates that structural information in assurance case graphs is highly predictive for link inference.

\begin{table}
\centering
\small
\setlength{\tabcolsep}{1.8pt}
\footnotesize
% \scriptsize
% \tiny
\caption{[RQ1] Performance comparison across three training and testing scenarios: Human (H) $\rightarrow$ Human, LLM (M) $\rightarrow$ H, and Mix (Human$+$LLM) $\rightarrow$ H. Metrics reported are ROC-AUC and F1 (mean $\pm$ std over 5 seeds).}
\label{tab:combined-all-scenarios}
\begin{tabular}{@{}lcc|cc|cc@{}}
\toprule
\multicolumn{1}{c}{\multirow{2}{*}{Model}} &
\multicolumn{2}{c|}{[RQ1a]~\textbf{H $\rightarrow$ H}} &
\multicolumn{2}{c|}{[RQ1b]~\textbf{M $\rightarrow$ H}} &
\multicolumn{2}{c}{[RQ1c]~\textbf{Mix $\rightarrow$ H}} \\
% & AUC & Prec. & Rec. & F1 & AUC & Prec. & Rec. & F1 & AUC & Prec. & Rec. & F1 \\
& \multicolumn{1}{c}{ROC-AUC} & \multicolumn{1}{c|}{F1} & \multicolumn{1}{c}{ROC-AUC} & \multicolumn{1}{c|}{F1} & \multicolumn{1}{c}{ROC-AUC} & \multicolumn{1}{c}{F1} \\
\midrule
\multicolumn{7}{c}{\textit{Decoder LLM (Text Only) In-Context-Learning}} \\
\midrule
Llama-3.2-1B & .509 $\pm$ .02 & .609 $\pm$ .02 & .485 $\pm$ .02 & \textbf{.591 $\pm$ .02} & .504 $\pm$ .02 & .566 $\pm$ .02 \\
Llama-3.1-8B & .577 $\pm$ .03 & .627 $\pm$ .01 & .552 $\pm$ .02 & .532 $\pm$ .01 & .559 $\pm$ .02 & .542 $\pm$ .02 \\
Qwen-2.5-7B & .588 $\pm$ .03 & .369 $\pm$ .07 & .592 $\pm$ .01 & .263 $\pm$ .07 & .570 $\pm$ .02 & .208 $\pm$ .06 \\
% \midrule
% \multicolumn{7}{c}{\textit{Encoder LM (Text Only)}} \\
% \midrule
% BERT-base & .677 $\pm$ .02 & .675 $\pm$ .05 & .581 $\pm$ .07 & .367 $\pm$ .19 & .661 $\pm$ .05 & .586 $\pm$ .09 \\
% BERT-large & .702 $\pm$ .03 & .651 $\pm$ .05 & .539 $\pm$ .05 & .342 $\pm$ .29 & .645 $\pm$ .04 & .552 $\pm$ .04 \\
\midrule
\multicolumn{7}{c}{\textit{Encoder LM+GNN}} \\
\midrule
GCN-1-layer & .687 $\pm$ .06 & .611 $\pm$ .13 & .490 $\pm$ .02 & .037 $\pm$ .06 & .599 $\pm$ .08 & .411 $\pm$ .11 \\
% GCNN-3 layers & .707 $\pm$ .03 & .702 $\pm$ .04 & .502 $\pm$ .01 & .020 $\pm$ .02 & .593 $\pm$ .07 & .414 $\pm$ .17 \\
SAGE-1-layer & .796 $\pm$ .03 & \textbf{.816 $\pm$ .02} & .636 $\pm$ .05 & \textbf{.525 $\pm$ .24} & .768 $\pm$ .02 & \textbf{.802 $\pm$ .02} \\
% SAGE-3 layers & .766 $\pm$ .05 & \textbf{.776} $\pm$ .07 & .540 $\pm$ .03 & .316 $\pm$ .11 & .718 $\pm$ .03 & .715 $\pm$ .05 \\
GAT-1-layer & .704 $\pm$ .05 & .634 $\pm$ .08 & .513 $\pm$ .02 & .201 $\pm$ .18 & .609 $\pm$ .08 & .455 $\pm$ .15 \\
% GAT-3 layers & .690 $\pm$ .05 & .605 $\pm$ .23 & .493 $\pm$ .02 & .043 $\pm$ .05 & .659 $\pm$ .06 & .601 $\pm$ .13 \\
\midrule
\multicolumn{7}{c}{\textit{MLM Self-Supervised+GNN}} \\
\midrule
% .855, .018, .717, .020,  .667, .011 , .448, .049,  .868, .019, .769, .016
UniGraph & \textbf{.855 $\pm$ .02} &  .717 $\pm$ .02 & \textbf{.667 $\pm$ .01} & .448 $\pm$ .05 & \textbf{.868 $\pm$ .02} & .769 $\pm$ .02 \\
% UniGraph-2 layer & \textbf{.848 $\pm$ .02} & .739 $\pm$ .02 & \textbf{.676 $\pm$ .02} & .524 $\pm$ .07 & \textbf{.879 $\pm$ .01} & .787 $\pm$ .01 \\
\bottomrule
\end{tabular}
\end{table}

\paragraph{RQ1b: LLM-generated data only Link Prediction}

% \input{tables/link_pred_r2}

% Table~\ref{tab:r2-synth2gold} 
Table~\ref{tab:combined-all-scenarios} reports the cross-domain performance where models are trained on LLM-generated assurance cases and evaluated on human-authored (gold) data. This setting assesses the model’s ability to transfer structural reasoning learned from LLM-generated data, potentially biased data, to authentic human reasoning structures. \textbf{When trained solely on LLM-generated assurance cases, GNN models transfer poorly to human-authored cases, showing significant drops in both AUC and F1}. All GNN variants' performance degraded in the LLM → Human scenario—for example, SAGE drops from (.796/.816) to (.636/.525), and GCN falls to (.490/.037). This indicates that distributional mismatch between LLM-generated and human-authored arguments hinders cross-provenance generalisation. In contrast, Decoder LLMs are not affected by LLM-generated data only and achieved better performance than GCN and GAT.

These results answer RQ2 by showing that LLM-generated data to human data on the link prediction task remains challenging, and models must explicitly address the reasoning bias introduced by LLM-generated data to achieve reliable human-level generalisation.

% \subsection{RQ3 (Human + LLM $\rightarrow$ Human): Semi-Supervised Link Prediction }
% \subsection{RQ1c: Semi-Supervised Link Prediction }
\paragraph{RQ1c: Semi-Supervised Link Prediction }

% \input{tables/link_pred_r3}

% Table~\ref{tab:r3-semi-sup} 
Table~\ref{tab:combined-all-scenarios} presents the results of the semi-supervised scenario, where models are trained jointly on both human-authored and LLM-generated assurance cases and evaluated on the gold human dataset. This setup examines whether incorporating LLM-generated data can improve generalisation and robustness in link prediction when gold data is limited. \textbf{The answer to RQ1c is: Yes, combining human and LLM-generated cases substantially boosts GNN performance, with UniGraph achieving the strongest generalisation across all models}. In the Mix → Human setting, nearly all GNN models improve over the LLM-only condition, and UniGraph reaches the highest overall scores (.868 AUC, .769 F1), demonstrating that incorporating diverse training sources mitigates distributional gaps and enhances model robustness.

\paragraph{RQ1: Ablation}

% We perform ablation by removing GNNs layer in UniGraph (w/o GNN) Fixed encoder and adding GNN depth on GCN, SAGE, GAT and UniGraph models for RQ1 on gold standard data.
% Table~\ref{tab:ablation-rq1} Fixed encoder results in a modest performance drop. indicating that structural reasoning remains effective even without encoder fine-tuning. 
% Across all evaluated GNN architectures (UniGraph, GAT, and GraphSAGE), increasing the number of graph layers does not improve and often degrades link prediction performance (Table~\ref{tab:depth-ablation}). This consistent trend suggests that deeper message passing is unnecessary for capturing the predominantly local dependencies in assurance case graphs.

To isolate the contribution of graph-based propagation, we compared the 1-layer UniGraph model against a text-only baseline~\ref{tab:ablation-rq1}. Removing the GNN layer resulted in a significant performance decrease, with ROC-AUC dropping from 0.855 to 0.677. This confirms that structural connectivity provides a critical signal for link prediction that cannot be captured by semantic text embeddings alone, directly addressing concerns regarding the motivation for a graph-based approach.

\begin{table}
\centering
\small
\setlength{\tabcolsep}{6pt}
\caption{Ablation study for RQ1 (Link Prediction) evaluating the contribution of graph propagation and GNN depth on Gold Standard.}
\label{tab:ablation-rq1}
\begin{tabular}{lcc}
\toprule
\textbf{Configuration} & \textbf{ROC-AUC} & \textbf{F1} \\
\midrule
UniGraph (1-layer GNN) & .855 $\pm$ .02 & .717 $\pm$ .02 \\
% \midrule
w/o GNN (Fixed Encoder Only) & .677 $\pm$ .02 & .675 $\pm$ .05 \\
% BERT-base & .677 $\pm$ .02 & .675 $\pm$ .05 & .581 $\pm$ .07 & .367 $\pm$ .19 & .661 $\pm$ .05 & .586 $\pm$ .09 \\

% UniGraph (1-layer GNN) & 0.796 $\pm$ 0.03 & 0.816 $\pm$ 0.02 \\
\bottomrule
\end{tabular}
\end{table}

% GCN-1 layer & .687 $\pm$ .06 & .611 $\pm$ .13 & .490 $\pm$ .02 & .037 $\pm$ .06 & .599 $\pm$ .08 & .411 $\pm$ .11 \\
% % GCNN-3 layers & .707 $\pm$ .03 & .702 $\pm$ .04 & .502 $\pm$ .01 & .020 $\pm$ .02 & .593 $\pm$ .07 & .414 $\pm$ .17 \\
% SAGE-1 layer & .796 $\pm$ .03 & \textbf{.816 $\pm$ .02} & .636 $\pm$ .05 & \textbf{.525 $\pm$ .24} & .768 $\pm$ .02 & \textbf{.802 $\pm$ .02} \\
% % SAGE-3 layers & .766 $\pm$ .05 & \textbf{.776} $\pm$ .07 & .540 $\pm$ .03 & .316 $\pm$ .11 & .718 $\pm$ .03 & .715 $\pm$ .05 \\
% GAT-1 layer & .704 $\pm$ .05 & .634 $\pm$ .08 & .513 $\pm$ .02 & .201 $\pm$ .18 & .609 $\pm$ .08 & .455 $\pm$ .15 \\
% % GAT-3 layers & .690 $\pm$ .05 & .605 $\pm$ .23 & .493 $\pm$ .02 & .043 $\pm$ .05 & .659 $\pm$ .06 & .601 $\pm$ .13 \\
\begin{table}
\centering
\small
\setlength{\tabcolsep}{6pt}
\caption{Effect of GNN depth on link prediction performance (RQ1). Shallow models AUC outperform deeper variants across architectures except GCN on the Gold Standard.}
\label{tab:depth-ablation}
\begin{tabular}{lcc}
\toprule
\textbf{Model} & \textbf{1-layer} & \textbf{3-layer} \\
\midrule
GCN & .687 $\pm$ .06 $\downarrow$ & .707 $\pm$ .03 $\uparrow$ \\
GraphSAGE  & .796 $\pm$ .03 $\uparrow$ & .766 $\pm$ .05 $\downarrow$ \\
GAT & .704 $\pm$ .05 $\uparrow$ & .690 $\pm$ .05 $\downarrow$ \\
UniGraph & .855 $\pm$ .02 $\uparrow$ & .833 $\pm$ .01 $\downarrow$ \\
\bottomrule
\end{tabular}
\end{table}

Furthermore, we evaluated the impact of GNN depth across architectures~\ref{tab:depth-ablation}. Results indicate that 1-layer models generally outperform 3-layer variants, with UniGraph showing a decrease in AUC from 0.855 to 0.833 at greater depths. This suggests that link inference in assurance cases relies primarily on local neighbourhood information; deeper architectures likely introduce noise or suffer from over-smoothing due to the hierarchical nature of argument graphs. GCN was the sole exception, showing improved performance with increased depth.

\begin{table}
% \footnotesize
% \small
\footnotesize
\centering
\caption{[RQ2a] Graph classification performance between Human vs LLMs assurance case to analyse bias in the data.
% RQ[3] The next two columns of Node and Edge show the explanation metrics results from the graph classification task.
}
\label{tab:graph-bias-exp}
\begin{tabular}{lcc}
\toprule
 % & \multicolumn{2}{c}{{[}RQ2{]}$\sim$Classify} \\
 % \midrule
\multicolumn{1}{c}{Model} & Acc & F1 \\
\midrule
\multicolumn{3}{c}{\textit{Decoder LLM (Text-Only) ICL}} \\
\midrule
% Llama3.2-1B-ICL-0 & .500 ± .037 & .071 ± .018 \\
% Llama3.2-1B-ICL-2 & .688 ± .301 & .230 ± .203 \\
Llama3.2-1B-ICL-4 & .608 ± .271 & .076 ± .069 \\
% Llama3.1-8B-ICL-0 & .086 ± .016 & .145 ± .020 \\
% Llama3.1-8B-ICL-2 & .403 ± .220 & .077 ± .050 \\
Llama3.1-8B-ICL-4 & .208 ± .098 & .178 ± .059 \\
% Qwen2.5-7B-ICL-0 & .156 ± .028 & .164 ± .028 \\
% Qwen2.5-7B-ICL-2 & .464 ± .114 & .119 ± .043 \\
Qwen2.5-7B-ICL-4 & .628 ± .094 & .120 ± .149 \\
\midrule
\multicolumn{3}{c}{Encoder LM + GNN} \\
\midrule
GCN-1 layer & .972 ± .00 & .984 ± .00 \\
% GCN-3 layers & .964 ± .02 & .979 ± .01 \\
SAGE-1 layer & .972 ± .00 & .984 ± .00 \\
% SAGE-3 layers & .972 ± .00 & .984 ± .00 \\
GAT-1 layer & .944 ± .05 & .969 ± .03 \\
% GAT-3 layers & .967 ± .01 & .980 ± .01 \\
\midrule
\multicolumn{3}{c}{\textit{MLM Self-Supervised+GNN}} \\
\midrule
UniGraph & .939 ± .03 & .964 ± .02 \\
\midrule
\multicolumn{3}{c}{\textit{Few-Shot Graph Cls (Encoder LM+GNN)}} \\
\midrule
GPrompt-GCN & .747 ± .21 & .834 ± .17 \\
GPrompt-SAGE & .831 ± .07 & .894 ± .05 \\ 
\bottomrule
\end{tabular}
\end{table}

\subsection{RQ2: Graph classification for Provenance Detection}
% \subsection{RQ2a: Graph classification Human vs LLMs}

\paragraph{RQ2a: Graph classification Human vs LLMs}

The results in Table~\ref{tab:graph-bias-exp} indicate that graph-based models can effectively differentiate between human-authored and LLM-generated assurance cases, achieving high accuracy and F1 scores. Models like GCN, GraphSAGE, and SAGE-3 layers reach near-perfect performance (Acc~$\approx$~0.97, F1~$\approx$~0.98), highlighting the presence of semantic biases between human and LLM-generated in the graphs. In contrast, GAT and Unigraph show slightly lower performance, suggesting that attention-based approaches may reduce sensitivity to superficial differences.

These findings reveal that LLM-generated assurance cases irregularities can be detected in the graph structure which diverges from typical in human-authored cases. This structural bias can serve as a diagnostic signal for improving structured arguments in generated data.

\begin{figure*}
    \centering
    \includegraphics[width=0.78\linewidth]{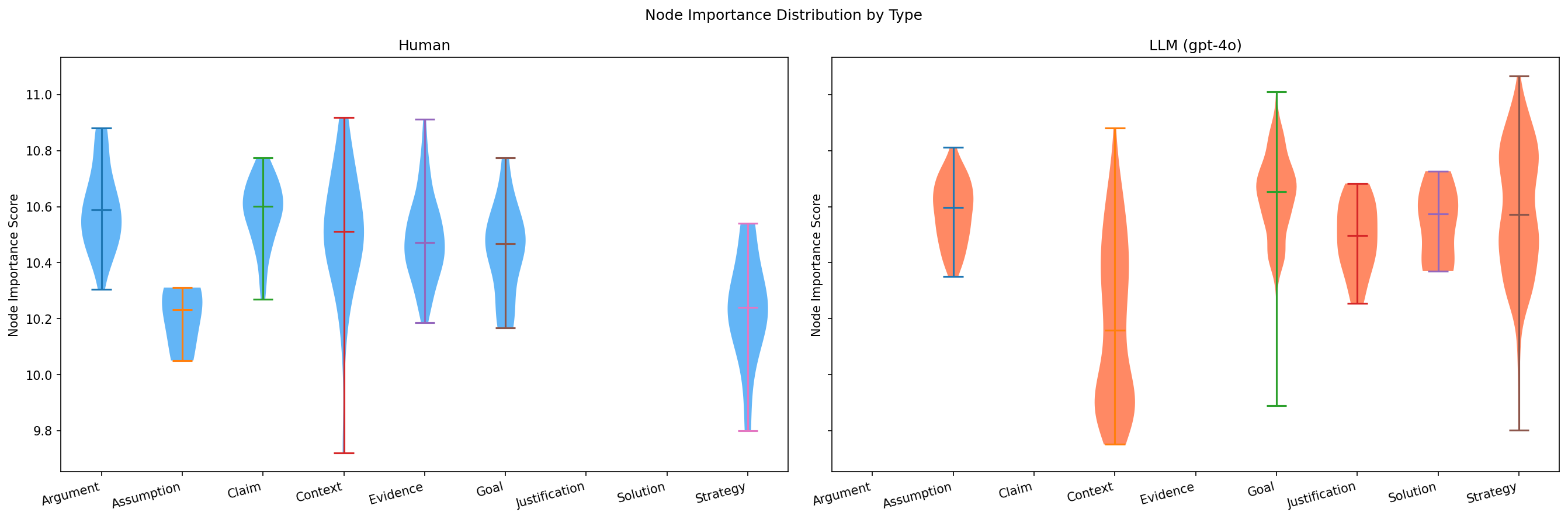}
    \caption{Node Importance Distribution by Type. Importance scores are obtained by GNNExplainer attributions from UniGraph model inference on the test set. Human-authored cases (blue) show a balanced distribution, while LLM-generated cases (orange) demonstrate high reliance on Goal/Strategy scaffolding and polarised Context integration.}
    \Description{Figure shows node Importance distribution by type. Importance scores are obtained by GNNExplainer attributions from UniGraph model inference on the test set. Human-authored cases (blue) show a balanced distribution, while LLM-generated cases (orange) demonstrate high reliance on Goal/Strategy scaffolding and polarised Context integration.}
    \label{fig:node-importance}
\end{figure*}

% Future work should explore causally constrained or counterfactual explanation techniques to ensure that high-performing discriminators also maintain high interpretability fidelity.

% \begin{figure}[ht]
%     \centering
%     \includegraphics[width=0.95\linewidth]{figures/GCN-graph_exp.pdf}
%     \caption{Selected example of GNNExplainer output from GCN model where the model has high positive fidelity and low unfaithfulness. Two colour bars indicate the intensity of attribution scores for the node and edge explanations.}
%     \label{fig:graph-example-gcn}
%     \Description{GNNExplainer visualisations for GCN models. Selected example of GNNExplainer output from the GCN model, where the model has high positive fidelity and low unfaithfulness. Two colour bars indicate the intensity of attribution scores for the node and edge explanations.}
% \end{figure}

\begin{figure}[ht]
    \centering
    \includegraphics[width=0.75\linewidth]{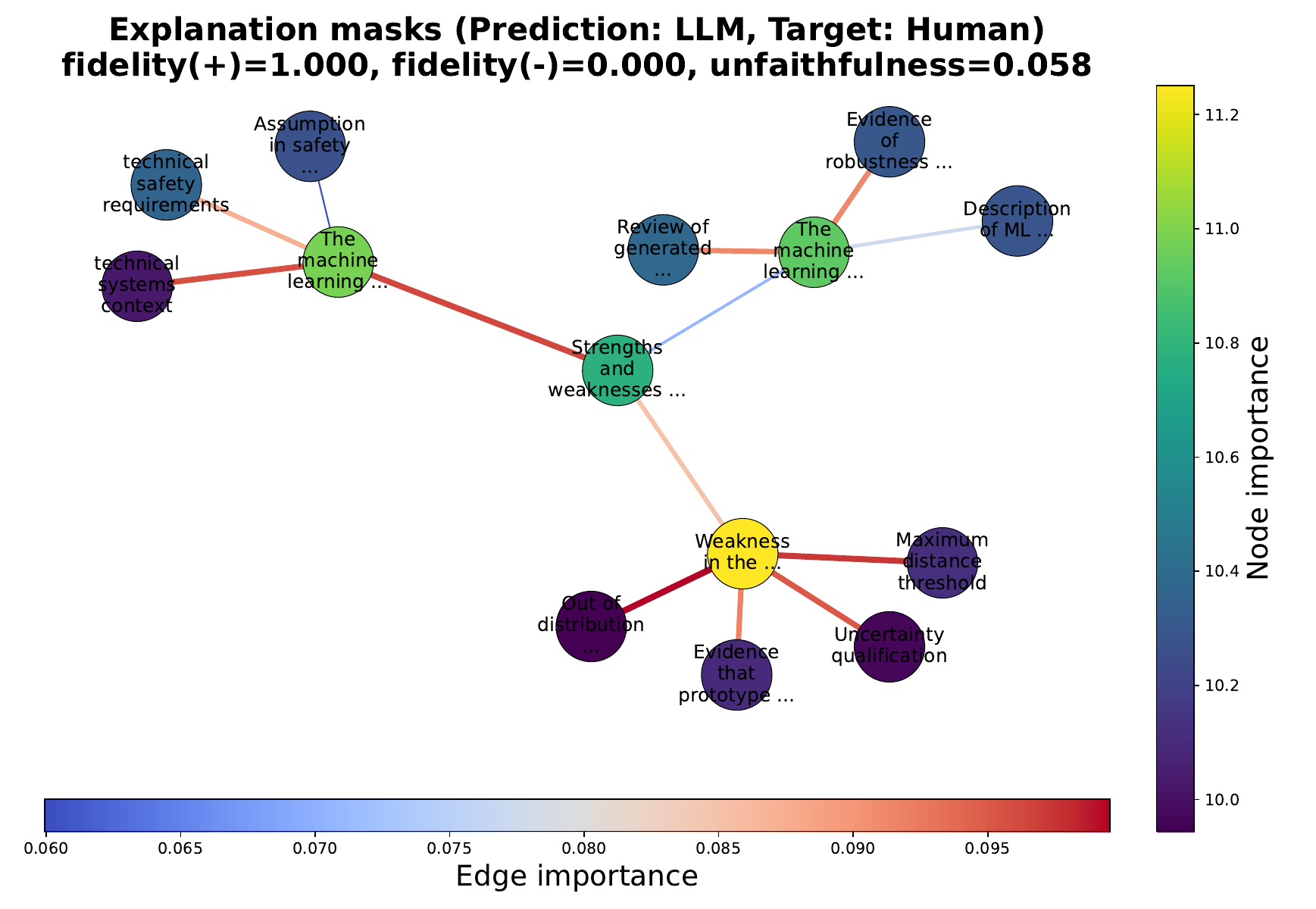}
    \caption{GNNExplainer output for UniGraph model showing high positive fidelity and low unfaithfulness with incorrect predictions. Colour bars indicate the intensity of attribution scores for the node and edge explanations.}
    \Description{GNNExplainer visualisations for GCN models. GNNExplainer output for UniGraph model showing high positive fidelity and low unfaithfulness with incorrect predictions. Two colour bars indicate the intensity of attribution scores for the node and edge explanations.}
    \label{fig:graph-example-uni}
\end{figure}

% \subsection{RQ2b: Human vs LLMs GNNExplainer Bias Analysis}
% \subsection{RQ2b: Exploratory Analysis: Explainability on Provenance Classification}
\paragraph{RQ2b: Exploratory Analysis of Explainability on Provenance Classification}

\begin{table}
\centering
% \small
\setlength{\tabcolsep}{2pt}
\footnotesize
% \scriptsize
\caption{[RQ2b] The GNNExplainer result of Node and Edge explanation metrics results from the graph classification task.}
\label{tab:graph-exp}
\begin{tabular}{@{}l|ccc|ccc@{}}
\toprule
\multicolumn{1}{c|}{{[}RQ2b{]}} & \multicolumn{3}{c|}{Node} & \multicolumn{3}{c}{Edge} \\ 
% & \multicolumn{3}{c}{[RQ3]~Node} & \multicolumn{3}{c}{[RQ3]~Edge} \\
\midrule
\multicolumn{1}{c|}{Model} & Fid$_{(+)} (\uparrow)$ & Fid$_{(-)} (\downarrow)$ & GEF($\downarrow$) & Fid$_{(+)} (\uparrow)$ & Fid$_{(-)} (\downarrow)$ & GEF ($\downarrow$) \\
\midrule
GCN-1 layer & .425 ± .33 & .550 ± .30 & .091 ± .02 & .017 ± .02 & .003 ± .01 & .015 ± .00 \\
% GCN-3 layers & .000 ± .00 & .142 ± .32 & .001 ± .00 & .036 ± .03 & .003 ± .01 & .018 ± .01 \\
SAGE-1 layer & .703 ± .40 & .128 ± .06 & .048 ± .02 & .011 ± .02 & .000 ± .00 & .015 ± .00 \\
% SAGE-3 layers & .000 ± .00 & .000 ± .00 & .000 ± .00 & .017 ± .02 & .006 ± .01 & .018 ± .00 \\
GAT-1 layer & .569 ± .44 & .431 ± .44 & .081 ± .03 & .031 ± .07 & .000 ± .00 & .012 ± .00 \\
% GAT-3 layers & .000 ± .00 & .194 ± .43 & .007 ± .01 & .028 ± .06 & .006 ± .01 & .013 ± .00 \\
\midrule
UniGraph & .567 ± .31 & .433 ± .31 & .114 ± .01 & .161 ± .14 & .044 ± .03 & .032 ± .02 \\
GPrompt-GCN & .431 ± .38 & .569 ± .38 & .057 ± .02 & .036 ± .06 & .025 ± .06 & .006 ± .01 \\
GPrompt-SAGE & .067 ± .08 & .708 ± .31 & .101 ± .06 & .092 ± .09 & .089 ± .10 & .036 ± .03 \\ 
\bottomrule
\end{tabular}
\end{table}

The faithfulness analysis evaluates whether the explanations provided by GNNs accurately capture the nodes and edges most influential to their predictions. The fidelity and GEF metrics in Table~\ref{tab:graph-bias-exp} reveal different results across models. Among all methods, GCN shows relatively balanced positive and negative node fidelity (Fid$_{(+)} \approx$~ 0.53, Fid$_{(-)} \approx$~ 0.47), suggesting moderate but interpretable attribution alignment with the model’s decision process. In contrast, SAGE-3 layers achieve near-zero or perfectly polarised fidelity values, reflecting overconfident decision boundaries where only a few nodes dominate the prediction. 

To assess the influence of specific argument elements, we analysed GNNExplainer importance scores categorised by GSN node type (Figure~\ref{fig:node-importance}). Human-authored data shows a distributed weighting across Argument, Claim, and Goal nodes. In contrast, LLM-generated data show a statistically skewed distribution, with disproportionate importance assigned to high-level Goal and Strategy elements. Furthermore, Context nodes in the LLM cohort show a bimodal distribution, suggesting inconsistent structural integration. These findings indicate that GNN provenance detection succeeds by identifying systematic hierarchical imbalances in LLM-generated data. Figure~\ref{fig:graph-example-uni} shows the node and importance edges of the  UniGraph model with high positive fidelity and low unfaithfulness; however incorrectly predicts the graph as LLM-generated.

% \begin{figure*}
%     \centering
%     \begin{subfigure}[]{0.40\textwidth}
%         \centering
%         \includegraphics[width=\linewidth]{figures/GCN-graph_exp.pdf}
%         \caption{GNNExplainer output for GCN model with high fidelity and low unfaithfulness with correct predictions for LLM-generated instance}
%         \label{fig:graph-example-gcn}
%     \end{subfigure}
%     % \hfill
%     \hspace{0.5cm}
%     \begin{subfigure}[]{0.38\textwidth}
%         \centering
%         \includegraphics[width=\linewidth]{figures/Unigraph-graph_exp.pdf}
%         \caption{GNNExplainer output for UniGraph model showing high positive fidelity and low unfaithfulness with incorrect predictions.}
%         \label{fig:graph-example-uni}
%     \end{subfigure}
%     \caption{GNNExplainer visualisations for GCN and UniGraph models. Two-coloured bars indicate the intensity of node and edge attribution scores.}
%     \label{fig:graph-example-sidebyside}
%     \Description{GNNExplainer visualisations for GCN and UniGraph models.}
% \end{figure*}

For edge-level fidelity, all models yield lower and more consistent scores (Fid$_{(+)} \approx$~ 0.04–0.07), indicating that link structures are less explicitly used for discrimination compared to node-level features. 

% Unigraph demonstrates the lowest GEF across both node and edge explanations, highlighting that its self-supervised graph embeddings produce more stable and less overfit explanations, albeit at the cost of slightly lower classification performance.

Overall, the faithfulness results indicate that while standard GNNs can effectively distinguish between human and LLM assurance cases, their explanations may not always reflect the true causal reasoning links that drive predictions. 

\section{Discussion}

Our findings demonstrate that, as a form of synthetic data, assurance cases generated by models like GPT-4o contain structural regularities and topological biases that differ from human-authored artefacts. While the specific nature of these structural signatures may vary across different proprietary or open-source LLMs, our framework demonstrates that GNNs can successfully detect these synthetic patterns. When using synthetic data as training data for the link prediction task, while GPT-4o-generated assurance cases offer a valuable pre-training signal, their structural biases limit transferability to real assurance data. GNNs trained solely on synthetic graphs internalise stylistic patterns that are different from hierarchical patterns created by humans. This results in poor generalisation in linking prediction tasks. UniGraph's robustness indicates that self-supervised objectives promote structure-aware learning regardless of specific graph topologies.

% Discussion.

The semi-supervised setting provides the best balance between generalisation and alignment with human reasoning. Models trained on both human and LLM-generated data show improved recall and robustness compared to those trained exclusively on one type. These results suggest that, despite their biases, LLM-generated assurance cases offer useful auxiliary supervision that aids GNNs in generalising reasoning link patterns in data-scarce situations. Specifically, GraphSAGE and UniGraph indicate that diverse linking distributions can enhance structural learning.

Previous studies, including the structured argument in a natural language inference model by \citet{ikhwantri2025explainable} and the traceability framework by \citet{etezadi2025classificationpromptingcasestudy}, mainly focused on predicting links between individual text elements. These approaches framed reasoning as a task of pairwise semantic similarity, aiming to identify whether a requirement should connect to a claim based on their textual embeddings. In contrast, this study focuses on global structural dependencies of a graph instance. We emphasise how argument structure arises from interactions within the assurance graph. This perspective allows for a deeper understanding of bias in reasoning arguments.

% The explainability analysis reveals that while GNNs achieve strong classification accuracy, their internal explanations vary widely in faithfulness. In several architectures, the explanations emphasise node features — such as the semantic embeddings of claims or arguments — more strongly than the relational links among them. This pattern suggests that the models often rely on semantic shortcuts rather than genuine reasoning chains when distinguishing between human and LLM-generated cases.

The explainability analysis reveals that, although Graph Neural Networks (GNNs) achieve high accuracy, their internal explanations often lack reliability. Many models prioritise node features from language embeddings over the edge between nodes, indicating reliance on semantic instead of structure when distinguishing between human and LLM-generated content. 

% Models such as GCN and SAGE contain relatively balanced node- and edge-level fidelity, indicating that they capture both the content and structure of the argumentation graph. However, models optimised for performance, like UniGraph, tend to prioritise node-level cues at the expense of edge interpretability, highlighting a trade-off between predictive accuracy and structural faithfulness.

SAGE often focus heavily on node-level cues, which biases provenance detection. However, UniGraph maintain a balance between node and edge fidelity, sacrificing edge interpretability, thus creating a trade-off between predictive accuracy and structural faithfulness, effectively capturing both content and structure. 

% This finding complements and extends prior text-based explainability studies. Earlier approach~\citep{ikhwantri2025explainable} focused on attention or gradient-based interpretations in language models, which identify salient tokens or sentences but cannot reveal whether reasoning decisions align with argument relations. Our graph-based approach introduces fidelity-based evaluation that measures whether model explanations correspond to the actual reasoning structure, moving from surface-level linguistic explanation to structural reasoning faithfulness.

These findings enhance previous studies on assurance case explainability. Past study~\citep{ikhwantri2025explainable} concentrated on text perturbation and gradient-based interpretations, which highlighted significant tokens but failed to clarify reasoning alignment with true causal relations. Our graph evaluation analyses a fidelity-based evaluation for bias detection of assurance case structure. 

In summary, our findings show that while GNNs can detect structural differences between human-authored and LLM-generated assurance cases, existing explanation methods show reliance on node features. Addressing this gap is an important avenue for future research.

% In summary, while current GNNs can differentiate between human and LLM-generated reasoning, their explanations may not truly reflect the underlying logic of assurance case development. Improving the faithfulness of these explanations is a vital area for future research in explainable graph reasoning for diagnosing compliance and safety for assurance cases.

\section{Threats to Validity}

% While the proposed framework demonstrates promising results for semantic and structural empirical analysis in assurance cases, certain threats to validity must be acknowledged:
\subsection{Construct Validity} Using link prediction and classification as proxies for argument quality may capture structural patterns without ensuring logical soundness. Additionally, the GNN might detect artefacts from the synthetic edge-generation prompt rather than intrinsic LLM reasoning styles. Finally, our findings are currently tied to GPT-4o-specific structural signatures and the limited faithfulness of existing GNN explanation tools.

\subsection{Internal Validity} The dataset used in this study, although carefully verified and curated to ensure the reliability and validity of human-authored samples, is relatively small. This limitation may affect the statistical power of our analyses and introduce potential biases that could impact the observed model performance. The restricted dataset size may also limit the ability of the models to generalise effectively across diverse assurance case structures. \textbf{However, publicly available human-authored assurance cases are scarce in practice in our setting, which reflects the situation whether in academia or industry due to data sensitivity}.

Our uncertainty analysis does not include variation over the LLM parameters. While this limits conclusions about full model variance, our evaluation reflects realistic deployment settings where LLM inference is performed via fixed APIs and variability from stochastic generation and prompt variation.

\subsection{External Validity} 
% The bias analysis conducted to differentiate between human-authored and LLM-generated assurance cases is limited to a one-way discrimination setup. While the graph classification models successfully distinguish between human and LLM-generated assurance case patterns, they do not address the iterative improvement of the LLM itself. Without a mechanism to provide feedback to the LLM generation process, the study's findings may not directly contribute to enhancing the quality of LLM-generated assurance cases. Future work is needed to explore approaches such as integrating the graph classifier as a discriminator within an iterative preference feedback loop or adversarial training framework. This method would enable the LLM to adapt its reasoning structures based on learned graph-level biases and improve the explainability and robustness of generated assurance cases.

The bias analysis conducted to differentiate between human-authored and LLM-generated assurance cases is limited to a one-way discrimination setup. While the graph classification models successfully distinguish between these patterns, the current framework does not yet involve an iterative improvement of the LLM itself. Consequently, while our study identifies structural biases, it does not directly evaluate how these insights might be used to enhance the generation process in real-time.

Furthermore, our empirical evaluation of LLM-generated assurance cases is currently limited to outputs produced by GPT-4o. We selected GPT-4o as a representative state-of-the-art model to establish the feasibility of our graph-based evaluation framework. Different LLMs (e.g., Claude, Gemini, or open-weight models) utilise different alignment training and generation heuristics. The specific topological biases (such as density or homophily) observed in our dataset may vary across models. \textbf{Conducting a large-scale, multiple LLMs empirical study to catalogue these variations falls outside the primary focus of this paper, as it would require extensive data generation and validation}. To address these practical constraints and mitigate the need for massive, multi-source training datasets, our framework leverages recent advancements in graph foundation models. Specifically UniGraph~\citep{10.1145/3690624.3709277} and GraphPrompt~\citep{10.1145/3543507.3583386}. These models require substantially less training data, which allows us to establish a robust proof-of-concept for graph-based assurance evaluation without relying on large-scale data generation.

These threats highlight areas where additional research and experimentation are needed to strengthen the generalisability and practical applicability of our evaluation framework and study.

\section{Conclusion and Future Work}

% Your submission of a finalised contribution for inclusion in the LREC Proceedings automatically assigns the above copyright to ELRA.

This study investigated how different modelling strategies perform across varying data provenance scenarios, focusing on the integration of masked language model text encoders and graph neural networks. 
Specifically, we explored five research questions. Our results show that graph neural networks can effectively learn reasoning and linkage patterns across two different types of assurance case structures. For RQ1, \textbf{link prediction} experiments demonstrated that GNNs perform well when trained on verified human-authored assurance cases (RQ1a). However, their performance declines in the LLMs-only data (RQ1b). It can transfer relational knowledge from the LLMs-generated data to real cases in a semi-supervised setting (RQ1c). This result highlights the structural and semantic gap between human and LLM reasoning.  

For RQ2a, \textbf{graph classification} analysis revealed that GNNs are capable of distinguishing human-authored from LLM-generated assurance cases by uncovering systematic structural biases. Finally, RQ2b showed that \textbf{explanation} fidelity remains limited, suggesting that current GNN explainers only partially capture the consistent reasoning paths of human vs LLMs assurance cases. Overall, these findings highlight the potential and challenges of graph-based reasoning in assurance cases.
% while underscoring the need for improved explainability and graph domain adaptation in future work.

Beyond the detection of structural biases, these findings establish a foundation for transitioning from static evaluation toward the iterative refinement of LLM-generated assurance cases. Future work could explore integrating our graph classifier as a discriminator within an iterative preference feedback loop or an adversarial training framework. 

%%
%% The acknowledgements section is defined using the "acks" environment
%% (and NOT an unnumbered section). This acknowledgement ensures the proper
\begin{acks}
% To Robert, for the bagels and explaining CMYK and colour spaces.
This work has been funded by the European Commission under grant agreement No. 101120606, CERTIFAI. This work has also benefited from the Experimental Infrastructure for Exploration of Exascale Computing (eX3), which is financially supported by the Research Council of Norway under contract 270053.
\end{acks}

\bibliographystyle{ACM-Reference-Format}
% \nocite{*}


% \newpage
% \appendix




\end{document}